\documentclass[12pt]{article}

\usepackage{scicite}
\usepackage{url}
\usepackage{times}
\usepackage{graphicx}

\newenvironment{sciabstract}

\title{A look into mirrors: A measurement of the $\beta$-asymmetry in $^{19}$Ne decay and searches for new physics}

\author{Dustin Combs$^{1,2}$, Gordon Jones$^{3}$, William Anderson$^{4}$, \\Frank Calaprice$^{5}$, Leendert Hayen$^{1,2}$, Albert Young$^{1,2}$\\
\\
\normalsize{$^{1}$Department of Physics,North Carolina State University}
\normalsize{Raleigh, NC 27695, USA}\\
\normalsize{$^{2}$Triangle Universities Nuclear Laboratory}
\normalsize{Durham, NC 27708, USA}\\
\normalsize{$^{3}$Department of Physics, Hamilton College} 
\normalsize{Clinton, NY , USA}\\
\normalsize{$^{4}$Johns Hopkins University, School of Medicine}
\normalsize{Baltimore, MD, USA }\\
\normalsize{$^{5}$Department of Physics, Princeton University}
\normalsize{Princeton, NJ, USA}
}

\date{}

\begin{document} 


\baselineskip24pt


\maketitle


\begin{sciabstract}


   High precision measurements of isospin $T=1/2$ decays in the neutron and nuclei provide strong model-independent constraints on extensions to the standard model of particle physics.  A measurement of the $\beta$-asymmetry in $^{19}$Ne decay between the initial nuclear spin and the direction of the emitted positron is presented which determined the zero intercept of the asymmetry parameter to be $A_0 = -0.03871(+65/-87)_{sys}(26)_{stat}$.  This result establishes $^{19}$Ne as the most precisely characterized nuclear mirror and fixes the Fermi-to-Gamow-Teller mixing ratio to $\rho = 1.6014(+21/-28)_{sys}(8)_{stat}$. The mixing ratio presented here is consistent with the previous, most precise measurement\cite{Calaprice75}, produces a value of the CKM unitarity parameter $V_{ud}$ in agreement with the nuclear mirror, neutron and superallowed $\beta$-decay data sets, shows no evidence for second class currents, and can be effectively used with neutron decay data to place a limits on exotic tensor couplings. 
\end{sciabstract}


\section*{Introduction}

Beta decay measurements provide precise and useful information concerning the weak interactions of quarks.  For these decays, the standard model (SM) of particle physics predicts vector (V) and axial-vector (A) couplings with a maximal parity violating, V-A Lorentz structure. We present a measurement (which is sensitive to this helicity structure) of the angular correlation  between the initial nuclear spin and the positron momentum, or $\beta$-asymmetry, in the isobaric analog decay of $^{19}$Ne to $^{19}$F. The results reinforce the usefulness of nuclear mirrors for beyond standard model (BSM) constraints, substantiate the current consistency of the data set for nuclear mirrors with the SM, address previous evidence for second class currents in $^{19}$Ne decays, provide constraints (when taken together with neutron decay data) on tensor couplings and highlight the possible impact for future measurements.

Data from beta decays are the basis, through tests of the unitarity of the Cabibbo-Kobayashi-Maskawa (CKM) mixing matrix and through constraints on exotic couplings, for some of our most stringent probes for BSM physics\cite{Gonzalez-Alonso:2019a,Holstein:2014a,NavBSM2013,Cirigliano:2013xha,gardner13a}. For couplings to the up quark, the expectation of unitarity in the SM is given by $\Sigma_{u} =|V_{ud}|^2 + |V_{us}|^2 + |V_{ub}|^2 = 1$, where $V_{ud}$, $V_{us}$ and $V_{ub}$ represent the strength of the coupling of the up quark to the down, strange and bottom quark, respectively. The experimental uncertainty for this test is 0.05\%, with the most precise experimental input coming from ``superallowed'' $0^{+}\rightarrow 0^{+}$ nuclear decays\cite{Hardy:2015a,PDG:2019a}. The $\Sigma_{u}$ unitarity test probes model-independent BSM interactions at energy scales up to 11 TeV for interactions with (V,A) symmetry, assuming there are no right-handed neutrinos\cite{Cirigliano:2009wk,Bhattacharya:2011qm}, while from the Large Hadron Collider (LHC) they are expected to reach 7 TeV. Recent improvements in the electroweak radiative corrections (EWRC) for neutron and nuclear decays by Seng {\it et al.}\cite{Seng:2018a,Seng:2019a} and cross-checked by Czarnecki {\it et al.}\cite{Czarnecki:2019a}, feature a reduced uncertainty for $V_{ud}$ (improving the reach of a unitarity test with $\Sigma _u$), but also of shift the central value of the EWRC by about 4 standard deviations from it's previously quoted value. The potential impact of this shift, about 0.1\% in the unitarity sum, motivates careful scrutiny of other aspects of the analysis of the superallowed data set, including the nuclear structure-based corrections\cite{Towner:2010bx}.  

Mirror decays in nuclei and the neutron offer a range of nuclear structure cases, permitting a complementary extraction of $|V_{ud}|$ to the superallowed data set\cite{NavSevHalfPlus2009,SevTownFtVals2008} and provide opportunities to optimize sensitivity for tests of specific BSM physics scenarios, for example exotic scalar (S) and tensor (T) couplings.  The isotope $^{19}$Ne has played an important role in fundamental symmetries studies since the 1950's\cite{Allen59,Calaprice69,Calaprice75,Adelberger:1983,Hallin:1985,Lienard:2015a}, in part because of the simplicity of its decay scheme and in part due to the sensitivity of some angular correlations measurements in this decay to the ratio of Gamow-Teller to Fermi amplitudes\cite{Calaprice75}. A current review of the physics impact of high precision decay data from $^{19}$Ne is presented in Rebeiro {\it et al.}\cite{Rebeiro:2019a}, with decay parameters included in Table I.  For a measurement of the positron distribution from polarized $^{19}$Ne decay, the leading order angular distribution\cite{Jackson1957} is given by 
\begin{equation}
  \label{decayrate}
\Gamma = 1 + \beta\left < P \right > A(W)cos\theta
 \end{equation}
with $\beta = v/c$ of the positron, $\left < P \right >$ the average $^{19}$Ne nuclear polarization, $\theta$ the angle between the nuclear spin and the positron momentum, and $A(W)$, the angular correlation parameter, determines the magnitude of the $\beta$-asymmetry as a function of relativistic energy, $W$. The $A$ parameter is specified to leading order for $^{19}$Ne by ${\bar A} \approx 0.67(\rho ^2  - 1.73\rho)/(1+\rho ^2))\approx -0.039$, with 
\begin{equation}
  \label{rho}
\rho \equiv g_A M_{GT}/g_V M _F \approx 1.60
\end{equation}
(using the sign convention of \cite{NaviliatCuncic:2009gi,Severijns:2008gw} for $\rho$, with $g_V$ and $g_A$ denoting the vector and axial vector weak coupling constants, and $M _{GT}$ and $M _{F}$ denoting the Gamow-Teller and Fermi matrix elements). The accidental cancellation which leads to the small value of $A(W)$ also leads to a very strong dependence of $A(W)$ on the $\rho$ parameter: $\delta {\bar A}/{\bar A} \approx -13\delta \rho /\rho$.  For a given relative precision level for $A$, this leads to over an order of magnitude higher relative precision for $\rho$, easing the requirements for the systematic uncertainties due to, for example, the $^{19}$Ne polarization.
\begin{figure}
	\centering
	\includegraphics[width= 0.95\columnwidth]{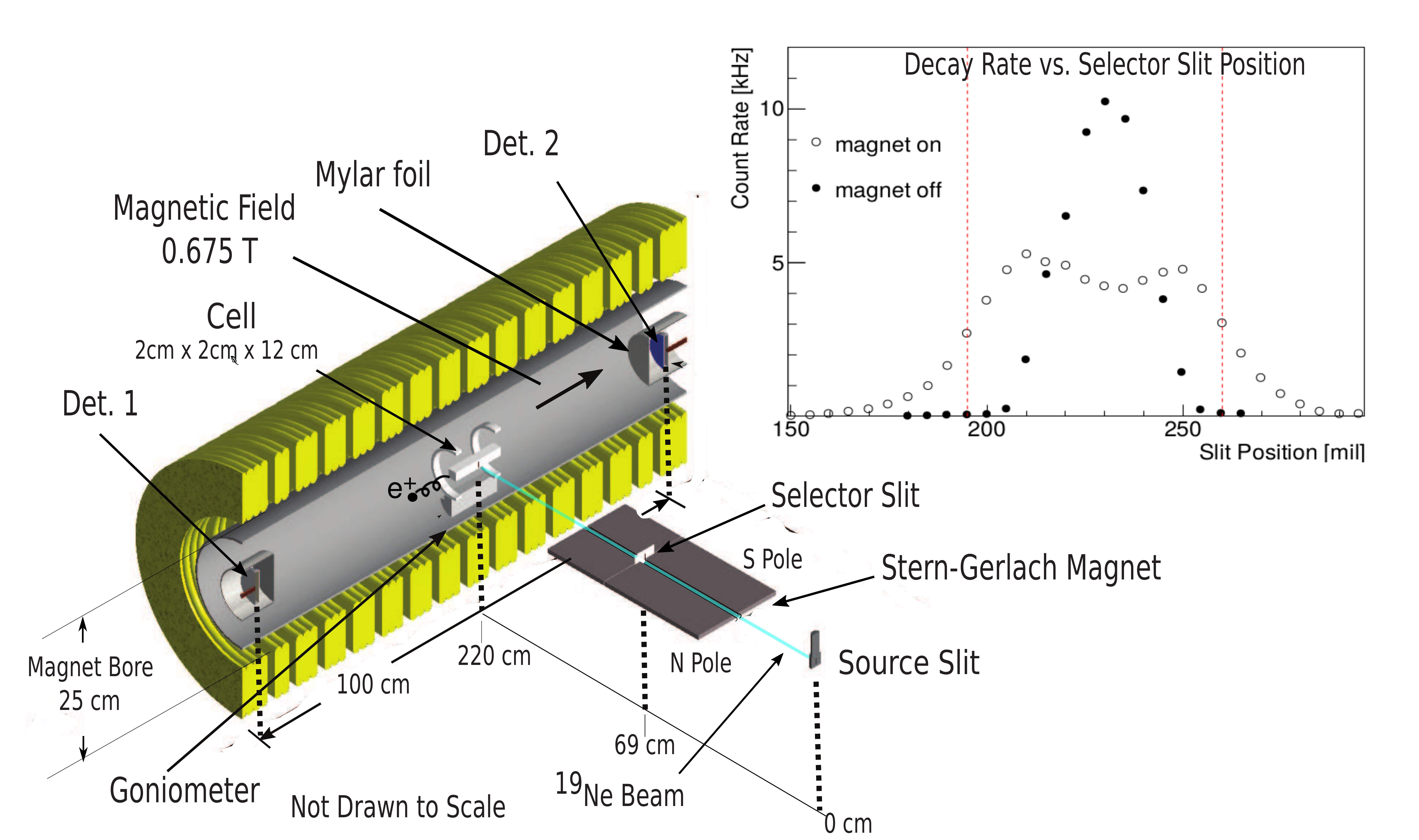}
        \caption{Schematic of the experimental apparatus.  Spin selection was accomplished using a Stern-Gerlach magnet and three slits: the atomic beam source slit, the movable selector slit and the entrance to the decay cell. The inset depicts a scan over the movable selector slit, with red vertical lines indicating the operational settings for each spin state and for the Stern-Gerlach magnet on and off.}
\label{fig:apparatus}
\end{figure}

\section*{Experimental Apparatus}
In 1995, an experiment to determine the $\beta$-asymmetry in the decay of $^{19}$Ne was performed at Princeton \cite{Jones:1996}.  Due to inconsistencies in the simulated and measured timing spectrum of scattered positrons, the results were not published immediately.  In this work we developed a new simulation and analysis of that experiment and a more detailed model of the detector signals which successfully reproduced the experimentally measured data.  We also incorporated a correction for $^{19}$Ne depolarization based on noble gas relaxation rates not available when our experimental data was obtained.

The $^{19}$Ne polarized atomic beam apparatus was developed at Berkeley\cite{Dobson:1963} and then moved to Princeton where it was used in a series of measurements until 1997\cite{Calaprice75,Calaprice:1985,Calaprice:1991}. The apparatus is described in detail in these publications, so we provide only a brief overview of the apparatus.  The $^{19}$Ne gas is created from the $^{19}$F(p,n)$^{19}$Ne reaction utilizing 12 MeV protons from the Princeton cyclotron in a flowing gas target, separated from the SF$_6$ target gas in a LN$_2$ trap, and pumped into a recirculation chamber containing a 2.5 cm$^3$ copper atomic beam source held at 38-40 K. For this work, a 1~in diffusion pump was added as the final compression stage for the oven, increasing the flux of $^{19}$Ne in the atomic beam by a factor of about 5 over previously reported work. 

The experiment after the $^{19}$Ne beam exits the ``source'' is depicted in Fig. \ref{fig:apparatus}.  The gas was polarized using a 44.8~cm long Stern-Gerlach magnet of the two-wire type described by Ramsey\cite{Ramsey:1956}.  The polarized atomic beam was defined by the three slits with vertical dimension of 1~cm: source (0.64~mm width), selector (0.71~mm width) and cell entrance (0.89~mm width).  The selector slit was positioned so that only atoms of a single spin state can enter the cell. Solenoidal magnetic fields positioned along the beam axis between the Stern-Gerlach magnet and the spectrometer magnet ensured that the polarization of the beam was preserved while traversing a differentially pumped buffer volume and the fringe fields of the spectrometer. Cell alignment was fine-tuned using a goniometer system to ensure optimum cell loading and polarizer conditions. A BGO scintillator detector was used to monitor the atomic beam source strength, to provide normalization for the backgrounds measured before and after the $^{19}$Ne decay runs.

The polarized gas entered the decay cell through a glass capillary array (Galileo Electro-Optics C25S10M10) into a 2~cm~$\times$~2~cm~$\times$~12~cm decay cell constructed of 0.5 $\mu$m thick mylar.  The mean residency time for $^{19}$Ne was 3.5~s. The decay trap sat inside a homogeneous, 0.675~T solenoidal magnetic field with its axis perpendicular to the atomic beam axis.  The magnetic field was manually shimmed to reduce inhomogeneities to less than 0.1\% over central 40~cm of the decay volume.  When the $^{19}$Ne decayed, positrons from the decay were guided by the magnetic field to a pair of lithium-drifted silicon (Si(Li)) detectors separated by a distance of 1.0 m.  The detectors were 7.46~cm diameter and 0.3~cm thick, with an active region 6.18 cm in diameter.  The detectors were divided into four quadrants, with each quadrant read out separately to reduce capacitance and rise time.  The detectors had a low energy threshold of 4~keV and resolution of about 2~keV for energy spectroscopy. 

\begin{figure}
   \includegraphics[width=0.95\textwidth]{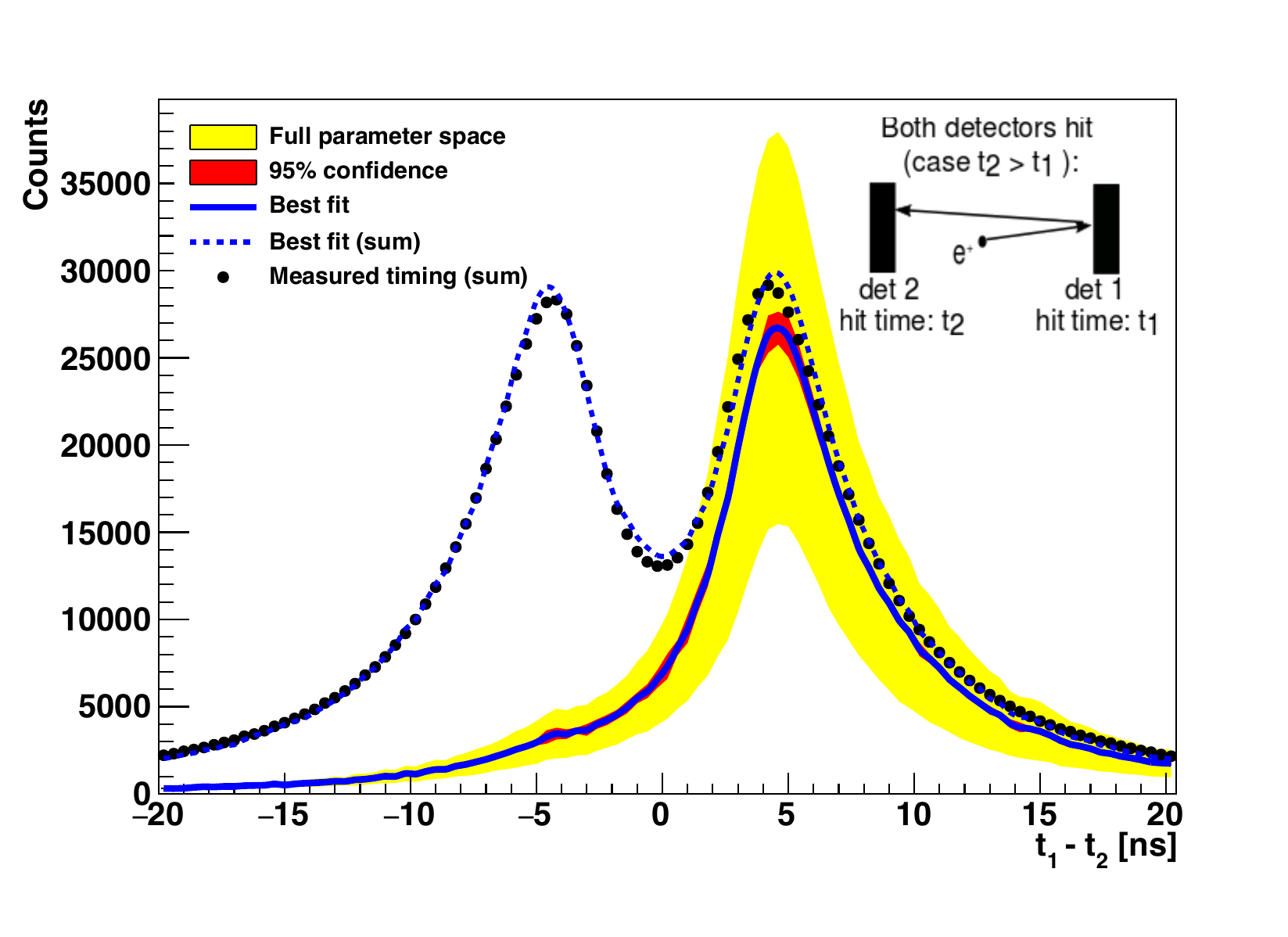}
   \caption{The measured timing distribution for backscattering events in which a positron hits both detectors (see inset).  The relative timing distribution is defined by $\Delta t = t_2 - t_1$, with $t_1$ and $t_2$ the earliest times in which a given positron hits the detectors.  Also shown is the best fit to the timing distribution with the Monte Carlo model of positron energy deposition and the 95\% confidence intervals for parameter variations in the timing response model.}
   \label{fig:timing}
\end{figure}

In this experimental geometry it is possible for a beta particle, initially emitted towards one of the detectors, to ``back-scatter'' from that detector and then to traverse the spectrometer and hit the opposite detector (see the inset in Fig. \ref{fig:timing}).  In these cases, the relative timing between the first ``hit'' on the two detectors was used to determine the initial emission direction.  This was accomplished by performing leading edge discrimination of the the detector pulses and then applying a ``walk-correction'' to account for the dependence of the timing on the amplitude of the recorded pulse.  Detector timing versus energy curves were measured in separate, dedicated runs using $^{60}$Co source with decays ``tagged'' by a fast plastic scintillator to provide a quantitative determination of the walk correction. Ultimately, knowledge of the polarization of the $^{19}$Ne and the corrections required for event timing reconstruction (backscattering) proved the sources of limiting uncertainty for this measurement.

Data collection was arranged in 8-section cycles with the spin states in the following order $\uparrow \downarrow \downarrow \uparrow \downarrow \uparrow \uparrow \downarrow$. A total of 38 hours of usable data comprised of approximately 6 million events were collected.  Data from the different spin cycles are first reconstructed by summing the energy over all hit quadrants and then using the relative timing to determine whether the intial decay was directed towards detector 1 or 2. After background subtraction, the resultant event rates in each detector were used to construct a super-ratio $R=\frac{N_{1\uparrow}N_{2\downarrow}}{N_{1\downarrow}N_{2\uparrow}}$, where the $N$ is the number of counts (corrected for background) and the subscripts refer to the detector number and spin state.  The ratio $R$ gives the asymmetry in each energy bin using the following relation $A_i=\frac{\sqrt{R_i}-1}{\sqrt{R_i}+1}$. This analysis allows for first order cancellation in systematic errors associated with differences in detector efficiencies and variations in the decay rate.  This ordering of the spin states provided additional suppression of drift in backgrounds and any residual dependence on drifts in detector thresholds, gain varations, {\it etc...}. 

Approximately 30\% of all events scatter from one detector into the other, leaving some amount of energy in both detectors. A timing spectrum is produced using the difference in the ``hit'' time of pulses from events leaving energy in both detectors: $\Delta t=t_1 - t_2$, where $t_1$ and $t_2$ are the earliest times the positron hits detector 1 and detctor 2, respectively.  The measured timing spectrum is depicted in Fig. \ref{fig:timing}.  Because the minimum transit time from one detector to another is 3.5~ns but charge collection times are of order 30~ns, the timing distributions for detectors 1 and 2 were not fully separated, and a simulation was developed to determine the correction required for mis-reconstructed events.

\section*{Monte Carlo Simulation}

The simulation assumes the decays occur uniformly throughout the interior volume of the holding cell, propagates positrons through the magnetic field using the approximation method of Bielajew \cite{Bielajew:2001jo} and incorporates interactions between the positrons and materials through the Monte Carlo calculations using the \textsc{penelope} code\cite{Salvat:1995}.  Energy pulses were created by propagating charge through the electric field within the detector assuming an ideal planar geometry\cite{Knoll:2000} and then through a fast amplifier with a 20~ns shaping time.  A separate simulation was performed for the {\it ex situ} $^{60}$Co timing data. All timing data from both the asymmetry runs in the spectrometer and $^{60}$Co data were fit to determine detector timing characteristics. This joint fitting procedure resulted in timing parameters which reproduced both data sets and provided the necessary predictions for incorrectly reconstructed backscattering events.

The measured timing spectrum is depicted in Fig. \ref{fig:timing}.  The analysis of the simulated timing data used five parameters: the detector energy threshold for timing signals, the electronic noise and a timing walk correction specified as a function of kinetic energy, $E$, as $w(E)=p_1+\frac{p_2}{p_3+E}$ \cite{Jones:1996,Spieler:1982}.  The parameters were allowed to vary in order to simultaneously fit the $^{60}$Co data and the positron coincidence timing spectrum from the asymmetry measurement using a $\chi ^2$ metric. The fit was strongly dominated by the constraints placed by the coincidence timing data from $^{19}$Ne decays, defining allowed ranges for the timing correction consistent with the measured data.  The results of this analysis are also depicted in Fig. \ref{fig:timing}. The fit indicates that 14(1)\% of the scattered decays or 4.1(3)\% of the total decays were incorrectly reconstructed. 

\begin{table}[htb]
\centering
\caption{List of corrections to the asymmetry and their uncertainties.  All values are multiples of $10^{-4}$.}
\begin{center}
\begin{tabular}{l c c}
\hline \hline
\textbf{Systematic} & \textbf{Correction ($10^{-4}$)} & \textbf{Uncertainty ($10^{-4}$)} \\ \hline
Monte Carlo Corrections: & & \\
\hskip 1 cm Above threshold in both detectors: & & \\
\hskip 2 cm Backscatter correction & -14.5 & $\pm$3.6\\
\hskip 2 cm Energy loss correction & +2.0 & $\pm$0.5 \\
\hskip 1 cm Above threshold in a single detector: & & \\
\hskip 2 cm Backscatter correction & -3.1 & $\pm$0.8\\
\hskip 2 cm Energy loss correction & +0.9 & $\pm$0.2 \\
\hskip 1 cm Below threshold in both detectors: & +0.5 & $\pm$0.1 \\ \hline

Polarization & -- & +0.0 -5.7\\
Spin relaxation & -5.3 & $\pm$5.3\\
Energy non-linearity & --  & $\pm$0.5 \\
Dead time & +0.5 & $\pm$0.4 \\
Pileup & -0.6 & $\pm$0.4\\ 
Background subtraction & -0.2 & $\pm$0.2 \\ \hline

\textbf{Total systematic} & \textbf{-19.8} & \textbf{+6.5} \textbf{-8.7} \\
\hline
Statistical & -- & $\pm$ 2.6\\ \hline
\textbf{Total} & \textbf{--} & \textbf{+7.0} \textbf{-9.1} \\
\hline \hline
\end{tabular}
\end{center}
\label{table:err}
\end{table}

\begin{figure}
   \centering
   \includegraphics[width=0.95\textwidth]{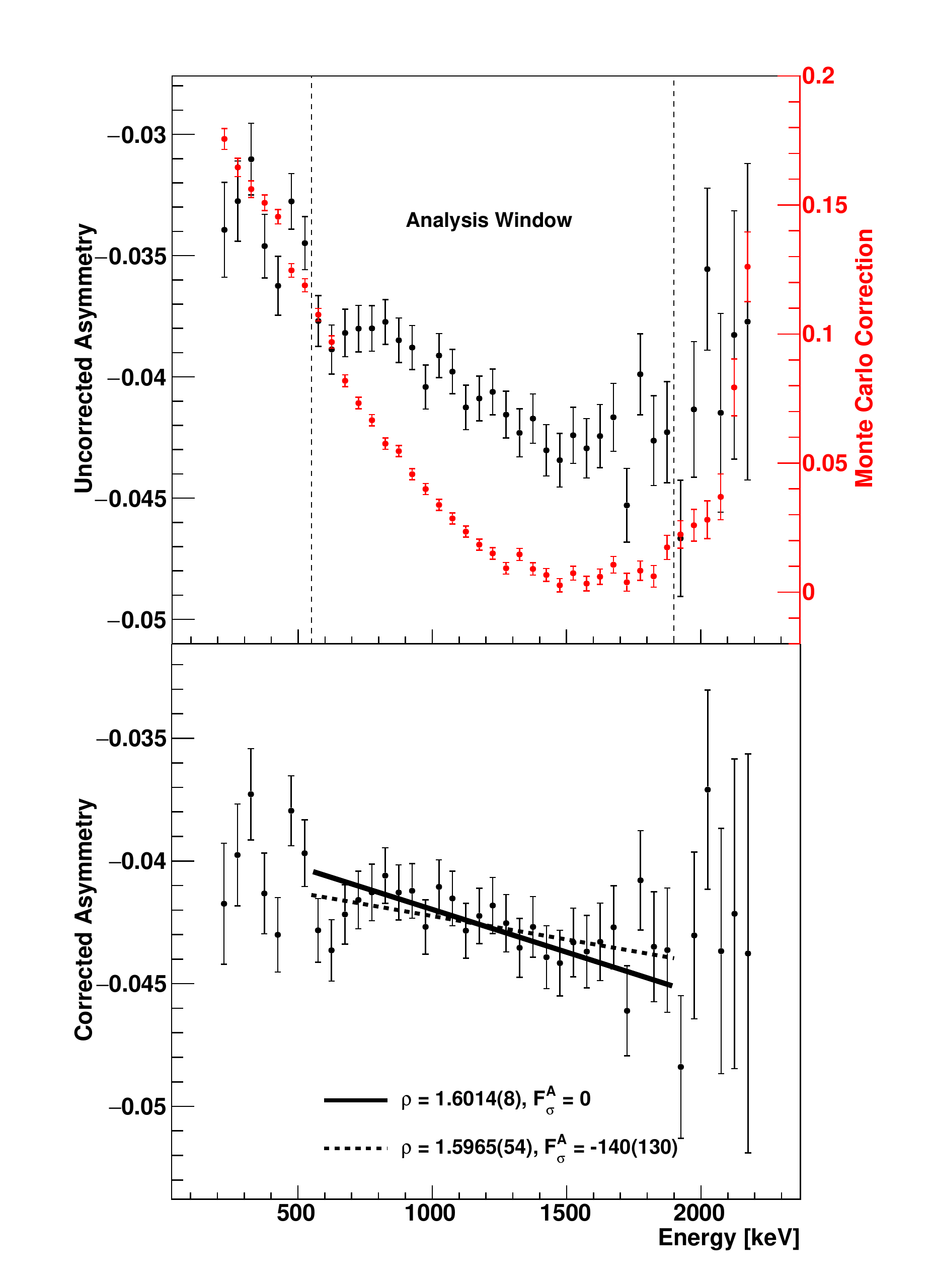}
   \caption{(top) The uncorrected asymmetry (statistical error bars) and the Monte Carlo correction to the magnitude of the asymmetry ($\frac{\delta A}{A}$) as a function of energy, (bottom) Corrected asymmetry with one and two parameter fit (bottom). In the one parameter fit, only $\rho$ is varied (as per the SM) and in the two parameter fit, both $\rho$ and the induced tensor, second class current coupling are varied.}
   \label{fig:asym}
\end{figure}

The raw asymmetry (no Monte-Carlo correction applied) as a function of energy is depicted in the top panel of Fig. \ref{fig:asym}, with our analysis window defined from 550~keV to 1900~keV to eliminate gamma backgrounds from positron annihilation and to minimize the magnitude of the corrections for positron scattering from cell components, the mylar isolation foils and the detector faces.  Corrections were determined using a Monte Carlo simulation of the detector geometry, depicted in the top panel of Fig. \ref{fig:asym}. 

\paragraph*{Results}
The $\beta$-asymmetry parameter as a function of relativistic energy, $A(W)$, is defined in Eq.~\ref{asyme} with input constants collected in Tab. \ref{avals}, based on expressions from Behrens and Buehring\cite{behrens1982electron}. In these expressions, we use an effective form-factor notation consistent with \cite{Holstein:1974cw}, where $F^V _0 = g_V M_{F}\stackrel{SM}{=}1$ and $F^A _0 = g_A M_{GT}$ are the leading order vector and axial-vector form-factors ($a$ and $-c$ in \cite{Holstein:1974cw}), and $F^V_{\sigma}$ and $F^A_{\sigma}$ are recoil order terms corresponding to the weak magnetism and induced tensor form factors ($-b$ and $-d$ in \cite{Holstein:1974cw}), with radiative corrections, some smaller terms due to recoil order matrix elements and radiative corrections included\cite{Hayen:2020a}. The expression below is accurate to an order of magnitude higher precision than required for the analysis of experiment. Further corrections depend on nuclear structure input and the experimental geometry, and are discussed elsewhere\cite{Hayen:2020a}. Because the vector coupling constants are very precisely specified and the induced tensor term is negligible in the SM ($F^A_{\sigma}$ is produced by second class currents), the only free parameter in the SM in Eq.~\ref{asyme} is $\rho = F^A _0/F^V _0$ (equivalent to Eq.~\ref{rho} but with the new notation).  When a one parameter fit for $A(W)$ as a function of $\rho$ is performed, the zero kinetic energy intercept is found to be $A_0 = -0.03871(+65/-87)_{sys}(26)_{stat}$ and $\rho = 1.6014(+21/-28)_{sys}(8)_{stat}$, in agreement with the most precise measurement to date of Calaprice {\it et al.}\cite{Calaprice75}. The $\chi ^2$/DOF for this fit was 1.12, with a probability of 31\% of measuring greater than this $\chi ^2$/DOF, indicating reasonable agreement with the SM functional form. The measured asymmetry and fit results are shown in the bottom half of Fig.~\ref{fig:asym}. 

\begin{equation}
  \label{asyme}
  A(W) = \frac{-\frac{2}{\sqrt{3}}\rho \left ( 1 + C_1 \right ) +  \frac{2}{3}\rho ^2\left ( 1  + C_2 \right ) + C_3}{1 + D_1 + \rho^2\left ( 1 + D_2 \right )} RC
\end{equation}

\begin{equation}
    C_1 = 10^{-5} \left (-287 + 3.21F^V_{\sigma} + 1.61F^A_{\sigma} + \left ( 61.0 - 0.65F^V_{\sigma} + 0.55F^{A}_{\sigma} \right ) W \right ) \label{c1} 
    \end{equation}
\begin{equation}
    C_2 = 10^{-5} \left (-335 + 6.42F^V_{\sigma} + 3.21F^A_{\sigma} + \left ( 79.9 - 2.41F^V_{\sigma} \right ) W \right ) \label{c2}
 \end{equation}
 \begin{equation}
    C_3 = 10^{-5} \left ( -1.10F^V_{\sigma} - 1.10F^{A}_{\sigma} \right )W \label{c3}
  \end{equation}
  \begin{equation}
    D_1 = 10^{-5} \left (-221 + 48.0W + 3.8/W - 2.5W^2 \right )  \label{d1}
  \end{equation}
  \[
    D_2 = 10^{-5} \left (-328 + 6.42F^V_{\sigma} +3.21F^{A}_{\sigma} + \left ( 77.9 - 2.41F^{V}_{\sigma} \right ) W + \right . 
  \]
  \begin{equation}
    \left . \left (-3.5 + 1.20F^{V}_{\sigma} - 0.60F^A_{\sigma} \right )/W - 4.3W^2 \right ) \label{d2}
\end{equation}
 \begin{equation}
  RC = 1 - 6.159\times10^{-4} +  2.640\times 10^{-3}/W  + 1.098\times10^{-4}~W - 9.572\times 10^{-6}~W^2
\end{equation}

\paragraph*{Corrections}
In order to quantify the fidelity of the Monte Carlo corrections to the asymmetry due to scattering effects, an assessment was made of the simulation results for events which hit more than one detector.  The reconstructed energy spectra and the fraction of events which undergo at least one, at least two and at least three or more scatters\cite{Jones:1996} were investigated, with the measured total hit fractions proving the most stringent test of the simulation.  These data, as well as more detailed investigations of $\beta$ scattering by the UCNA collaboration\cite{Martin:2003a, Martin:2006a,plaster12}, indicate a relative uncertainty of about 25\% is a conservative and appropriate estimate for our Monte Carlo corrections to the asymmetry with the largest relative discrepancy between simulation and measurement applied to the ratio of (two or more scatters)/(three or more scatters) of 21\%.

Table \ref{table:err} lists the systematic corrections to the asymmetry and their associated uncertainties.  For the scattering corrections, each correction listed in the table is the difference between the uncorrected experimental $A_0$ and the value found after applying the Monte Carlo correction for that class of event and then performing a fit to the asymmetry to extract $A_0$.  In what follows we provide a brief description of our evaluation of the leading sources of systematic corrections and uncertainty.

The Stern-Gerlach magnet provided a uniform 24 kG/cm gradient over the entire beam height.  For $^{19}$Ne at a temperature of 38 K, this gradient causes a typical deflection of 640 $\mu$m over the length of the magnet, with 97\% of the beam being deflected by more than 250 $\mu$m.  The selection slit can block the line of sight between the source and cell, allowing only atoms whose trajectories are bent by the magnet to enter the cell.  Larger slits before and after the magnet, and before the solenoid, are used to reduce the number of unpolarized $^{19}$Ne atoms that diffuse into the solenoid chamber.  The inset in Fig. \ref{fig:apparatus} shows the count rate in the decay cell as a function of the selection slit position with the magnet on and off.  Vertical lines show the positions used during the experiment(4.93~mm and 6.58~mm). The glass channels of the MCP cell entrance slit was slightly rotated with respect to the beam, causing a slight asymmetry in the beam profile with the magnet on.

To determine the uncertainty in the polarization, the background-subtracted, beta detection rate was measured with the magnet on and off (see the inset in Fig. \ref{fig:apparatus}). For spin-down selection at slit position 6.58 mm, the magnet off rate is 2.2\% of the magnet on rate.  For spin-up selection at slit position 4.93 mm, the magnet off rate falls to 0.7\% of the magnet on rate.  In order to set a lower limit on the polarization, it is assumed that the unpolarized atoms detected with magnet off are still present with the magnet on, causing on average 1.5\% of the beam to be unpolarized.  This is a conservative limit because atoms with the wrong spin are actually deflected away from the slit with the magnet on, even if there is a line of sight between source and cell.  A simple model of the beam found that there should be no spin contamination in the cell by a 0.25 mm margin. In practice, any background would far more likely be of the correct spin state as most of the wrong spin state is blocked by a differential pumping slit well upstream of the entrance slit.

The depolarization rate was not measured for the cell used in this experiment, however stringent limits on $^{19}$Ne depolarization were determined through measurements performed by Schreiber\cite{Schreiber:1983}.  They used the same atomic beams machine, solenoidal magnetic field geometry and a very similar cell (with an MCP entrance channel, roughly the same dimensions and plastic with the same elemental constituents), with no $^{19}$Ne relaxation observed.  These measurements, scaling for the difference in expected wall collision rates and holding times due to cell geometry, place an upper limit of 2.3\% (95\% C.L.) for spin depolarization in the cell.  During the time since this experiment and the first analysis were complete, a relaxation time of 900~s was measured for $^3$He on mylar by Heil \cite{heil2017}, which we scale for $^{19}$Ne in our geometry assuming dipole-dipole relaxation in the wall material dominates the depolarization\cite{jacob2003fundamental,cain1990nuclear} to obtain a correction of 1.4\%. Given the possibility of relaxation at this level, we apply a correction to the asymmetry of 1.4(1.4)~\% accounting for depolarization over a larger range than covered by the relaxation limits from the measurements of Schreiber.
 
The linearity of the detector's energy response was checked with sources $^{241}$Am, $^{133}$Ba, $^{60}$Co and $^{207}$Ba.  The maximum non-linearity was measured to be 1\% at 1550 keV, the Compton edge of the $^{207}$Bi gamma line.  The asymmetry is very insensitive to linearity errors, leading to an upper limit for the uncertainty in $A_0$ due to the non-linearity of 0.14\%.  Backgrounds in the beta signals are determined by moving a brass flag to cover the entrance slit to the cell, with the background dominated by polarized $^{19}$Ne decaying outside the cell but also with a smaller unpolarized component due to ambient background.  The ratio of signal to background in the analysis window is greater than 100 when determined in this way, with a 6\% correction due to the component of the atomic beam which actually enters the cell when the brass flag is not present.  The resultant correction to the asymmetry is found to be 0.2(2)\%.

\paragraph*{Two Parameter Fit}
\renewcommand{\tabcolsep}{0pt}
\begin{table}
\caption{Parameters used to calculate $\mathcal{F}t_0$ and $A_0$.}
\begin{center}
\begin{tabular}{l c c c}
\textbf{Constant} & \textbf{Value} & \textbf{Units} & \textbf{Reference} \\
$K/(\hbar c)^6$ & $8120.278(4) \times 10^{-10}$ & \, GeV$^{-4}$s & \cite{Amsler:2008kq}\\
$G_F/(\hbar c)^3$ & $1.16637(1)\times 10^{-5}$ & GeV$^{-2}$ & \cite{Amsler:2008kq}\\
$\Delta^V_R$ & $2.467(22)\times 10^{-2}$ & & \cite{Seng:2018a} \\
$Q_{EC}$ & $3.23949(16)$ & MeV & \cite{ame2016}\\
$f_V$ & $98.648(31)$ & &\cite{towner2015parametrization} \\ 
$f_A/f_V$ & $1.0012(2)$ & & \cite{Hayen2020c} \\ 
$\delta^V_{C}-\delta^V_{NS}$ & $0.52(4)\times 10^{-2}$& & \cite{Severijns:2008gw}\\
$BR$  & $0.999878(7)$ & \ & \cite{Rebeiro:2019a}\\
$P_{EC}$ & $0.00101(3)$ & \ &\cite{Severijns:2008gw, Bambynek1977}\\
$t_{1/2}$ & $17.2578(34)$ & s & 
\cite{Broussard:2014a,Fontbonne:2017mwx} \\ 
$F^V_{\sigma}$ & $148.5605(26)$ &  & \cite{NaviliatCuncic:2009gi,Stone:2005ez}\\
$W_0$ & $2.72850(16)$ & MeV & \cite{ame2016} \\ 
$m_e$ & $0.510998910(13)$ & MeV & \cite{Mohr:2012dr}\\
$M$ &  $19.00188090(17)$ & amu & \cite{ame2016} \\
\end{tabular}
\end{center}
\label{avals}
\end{table}

Analysis of the energy dependence of the asymmetry in ref. \cite{Calaprice75} determined a non-zero value for $F^A_{\sigma} = 250(100)$. The primary motivation for using Si(Li) detectors in this work was to permit more reliable modeling of the energy dependent response than previous experiments performed with plastic scintillators. A two parameter fit was performed, in which $\rho$ and $F^{A}_{\sigma}$ were both allowed to vary, with the variation in $\chi ^2$ used to determine the 67\% confidence level assuming Gaussian statistics. The results of the analysis are $\rho = 1.5965(+21/-28)_{sys}(54)_{stat}$ and  $F^{A}_{\sigma} = -140(+43/-29)_{sys}(130)_{stat}$, in nominal agreement with the SM and the one parameter fit,  and in disagreement with Ref. \cite{Calaprice75} which measured a slope with opposite sign. The uncertainty in the slope is dominated by uncertainties in the Monte Carlo corrections to the asymmetry. We find no evidence for second class currents. We identify no obvious origin for the disagreement with Ref. \cite{Calaprice75}, but the small magnitude of the slope makes measurements of the energy dependence very sensitive to the detector response function (linearity, scattering contributions, possible dead layers) and backgrounds.  With the improved signal-to-background ratio and the ability to reliably model the energy response for our measurement, we believe it represents an improvement in the understanding of the asymmetry's energy dependence.

\section*{Discussion}
\begin{figure}
   \centering
   \includegraphics[width=0.95\textwidth]{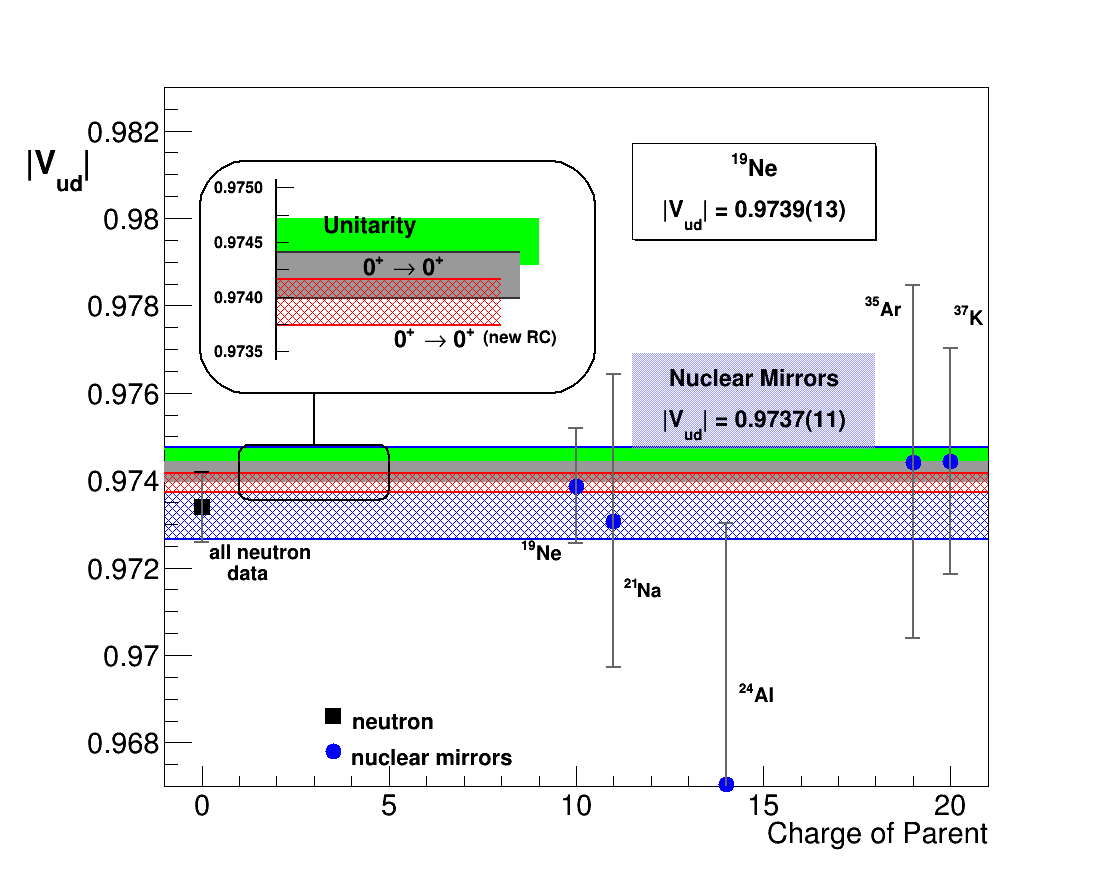}
   \caption{A global summary of $|V_{ud}|$ determined from nuclear decays. Mirror decay values taken from \cite{Hayen2020c} except for $^{24}$Al\cite{adelberger1985} (technically a $T=1$ isotriplet decay) and $^{19}$Ne (this work). Also presented are $|V_{us}| = 0.2243(9)$ from Ref. \cite{PDG:2019a,Czarnecki:2019a} for the ``unitarity'' band, $|V_{ud}|(0^+ \rightarrow 0^+)=0.97420(21)$ from Ref. \cite{Hardy:2015a} with the previous vertex correction values of Marciano and Sirlin\cite{Marciano:2005ec}, $|V_{ud}|(0^+\rightarrow 0^+) = 0.97395(21)$ with the new nuclear and EWRC corrections of Seng {\it et al.}\cite{Seng:2018a,Seng:2019a} and $|V_{ud}|=0.9734(8)$ for the neutron with the new EWRC (see text).}  
   \label{fig:Vudsummary}
\end{figure}

To compare to other nuclei, the quantity $\mathcal{F}t_0$ is defined in Naviliat-Cuncic and Severijns\cite{NaviliatCuncic:2009gi},
\begin{equation}
\label{Ft0}
 \mathcal{F}t_0 \equiv f_V t_{1/2}\left (\frac{1 + P_{EC}}{BR} \right)\left ( 1 + \delta _{NC} - \delta _{C}\ \right )\left (1 + \delta _R ' \right )\left [1 + \left ( \frac{f_A}{f_V} \right )^2 \rho ^2 \right] 
\end{equation}
using quantities from Tab. \ref{avals}. The mean value and uncertainty in the lifetime of $^{19}$Ne was calculated from the error weighted average of all published experimental results 
\cite{Broussard:2014a,Fontbonne:2017mwx}
with a scale factor of 1.9 for scatter. The value of $f_V$ was calculated using the parameterization of Towner and Hardy \cite{towner2015parametrization} with the $Q_{EC}$ value of Ref. \cite{ame2016}. We find that $\mathcal{F}t_0 = 6142(17)$~s for $^{19}$Ne, dominated by the uncertainty in $\rho$.  From this value of $\mathcal{F}t_0$, 
\begin{equation}
\label{Vud}
  \mathcal{F}t_0 = \left [ \frac{K}{G_F ^2 |V_{ud}| ^2 \left (1 + \Delta ^V  _R \right )} \right ]
 \end{equation}
the EWRC of Seng {\it et al}\cite{Seng:2018a} and the new analysis of Gamow-Teller decays of Hayen\cite{Hayen:2019a, Hayen2020c}, we can extract the value of $|V_{ud}|$ from $^{19}$Ne decay, 
$|V_{ud}|$($^{19}\mathrm{Ne})= 0.9739(13)$ and compare it to the value from other decays, summarized in Fig. \ref{fig:Vudsummary}, finding good agreement with the other mirror decays, the neutron and the superallowed data set.  For neutron data we used all published values and particle data group methods to determine averages and global uncertainties, obtaining a neutron lifetime of $879.8(7)$~s, with scale factor of 2.0 for scatter and $g_A = 1.2756(10)$, with a scale factor of 2.2 for scatter, resulting in $\mathcal{F}t_0$~(neutron)~$= 6148(10)$~s. Our result does not include recently discovered contributions from quasi-elastic processes to the EWRC \cite{Gorchtein2018}, which are expected to contribute at the $1 \times 10 ^{-4}$ level. 

The value of $\mathcal{F}t _0$ is sensitive to exotic tensor couplings, primarily through Fierz interference terms that impact the measured decay rate, $\beta$ spectra, and angular correlations\cite{Profumo:2007a,Bhattacharya:2011qm,Gonzalez-Alonso:2016a}. Beta decays are competitive with the LHC for left-handed neutrino couplings\cite{Gonzalez-Alonso:2019a}, with a particularly sensitive constraint for these produced from the ratio of $|V_{ud}|^2$ determined from the neutron to that from $^{19}$Ne, $\mathcal{R} =  |V_{ud}|^2$(neutron)/$|V_{ud}|^2(^{19}$Ne$)$\cite{Bhattacharya:2011qm}.  To illustrate the impact of this constraint, for model independent analysis, one can take $C_S = (G_F/\sqrt{2})V_{ud}g_S \epsilon _S$ and $C_T = (G_F/\sqrt{2})V_{ud}g_T \epsilon _T$, with the $g_S = 1.02(10)$, and $g_T = 0.989(34)$ the isovector scalar and tensor charges (or form factors) of the nucleon and $\epsilon _S$ and $\epsilon _T$ are effective BSM scalar and tensor couplings\cite{Gupta:2018a}, giving $\mathcal{R}\approx 1 + 0.51 \epsilon _S - 6.1\epsilon _T = 0.9990(52)\ (90\% \mathrm{C.L.})$. In this expression, the current limits on scalar couplings are taken from the superallowed data set\cite{Hardy:2015a} and the impact of the Fierz term on asymmetry measurements is assumed to be an average dilution factor, consistent with the analysis procedures of the three most recent neutron $\beta$-asymmetry experiments which perform single parameter fits of the asymmetry to determine $\rho$\cite{Gonzalez-Alonso:2016a,Pattie:2013a,Pattie:2013b}.  This analysis yields $\epsilon _{T} = 2.8(8.7)\times 10^{-4}$ (at 90\% C.L.), which corresponds to an energy scale for new tensor interactions of $\Lambda _T \equiv (174$~GeV$)/\sqrt{\epsilon _T} > 5.1$~TeV.  Details of this analysis can be found in Ref.~\cite{Hayen:2020a}.

\begin{table}
\caption{Summary of results.}
\begin{center}
\begin{tabular}{l c}
\textbf{Observable} & \textbf{Value} \\
$A_0$ (SM) & $-0.03871(+65/-87)_{sys}(26)_{stat}$ \\
$\rho$ (SM: $F^A_{\sigma}$ = 0) & $1.6014(+21/-28)_{sys}(8)_{stat}$\\
$\rho$ (BSM: $F^A_{\sigma} \neq$ 0) &  $1.5965(+21/-28)_{sys}(54)_{stat}$\\
$F^A_{\sigma}$ & $-140(+43/-29)_{sys}(130)_{stat}$\\
$ft$ & $1721.6(1.0)$~s\\
$\mathcal{F}t_0$ &  $6142(17)$~s\\
$V_{ud}$ & $0.9739(13)$ \\
$\Lambda _T \equiv (174$~GeV$)/\sqrt{\epsilon _T}$ & $> 5.1$~TeV
\end{tabular}
\end{center}
\label{summary}
\end{table}

Exotic coupling limits such as these are the natural output of global fits to the beta decay data set. The sensitivity quoted here is less strong than the global fit to all beta decay data by Wauters {\it et al.},  with a ``sensitivity scale'' determined by the precision of the constraints of about 7.8 TeV, and comparable to that of Gonzalez-Alonso {\it et al.}\cite{Gonzalez-Alonso:2019a}. We did, however, incorporate more recent neutron decay data than these earlier publications which should improve their quoted limits somewhat.  Current, model independent limits for tensor couplings from the LHC are also about 7.8 TeV\cite{Gupta:2018a}, which are expected to improve to about 11 TeV when the full LHC data set is analyzed.

If it is possible to push the precision of the $\mathcal{F}t_0$ values for the neutron and $^{19}$Ne to levels comparable to the superallowed decays (~1s), one can establish a model-independent sensitivity scale with ``discovery potential'' for BSM tensor couplings\cite{plasterva2013} above 20 TeV.  These measurements would also probe uncertainty scales well below the recent shifts in the EWRC, relevant to both the uncertainties in the isospin-mixing nuclear-structure corrections for the superallowed decays\cite{Towner:2010bx} and proposed new structure corrections to the EWRC\cite{Seng:2019a}. These shifts have sharpened motivation to develop a systematic nuclear structure analysis of nuclear beta decay over a wide range of nuclei, an activity already underway with promising approaches in the mass range of $10 < A < 20$ using effective field theory approaches\cite{Gysbers2019, Barrett2013}. From an experimental perspective, all of the ingredients for this improvement in precision are in place.  For the neutron, the next generation of measurements are planned at the sensitivity level of the superallowed decays.  For $^{19}$Ne, using optical trapping methods, Fenker {\it et al.}\cite{Fenker:2016a,Fenker:2017a} has already demonstrated it is possible to measure the $\beta$-asymmetry with systematic errors an order of magnitude smaller than those presented here, at about the 0.3\% level, with improvements to 0.1\% precision underway\cite{Melconian:2019a}.  Such a measurement of $^{19}$Ne would involve a significant focus on the Neon isotopes, but exactly such an experimental program is underway at the Hebrew University\cite{Mardor:2018a}. 

The results in this article are summarized in Table 3, and establish $^{19}$Ne decay as the most precise asymmetry measurement from a nuclear mirror.  The value of the zero-intercept of the asymmetry was determined to be $A_0 = -0.03871(+65/-87)_{sys}(26)_{stat}$, with the energy dependence of the asymmetry showing no evidence for second class currents, disagreeing with the results of ref. \cite{Calaprice75}. When taken together with recent theoretical progress, the nuclear mirror data set for $|V_{ud}|$ is internally consistent, and also consistent with the neutron and superallowed decays.  The sensitivity of $\beta$-asymmetry measurements to the mixing ratio $\rho$, when taken together with all other details of the decay already being known to roughly two parts in $10^4$, makes this a unique target of opportunity for fundamental symmetries studies. State-of-the-art experimental technique could already, in principle, probe precision levels comparable to the super-allowed decays. The ratio of the $\mathcal{F}t _0$ value for $^{19}$Ne with that from the neutron provides strong, model-independent constraints for BSM tensor couplings and opens a path to unprecedented sensitivity.



\bibliography{scibib_v2}

\begin{thebibliography}{10}

\bibitem{Calaprice75}
F.~P. Calaprice, S.~J. Freedman, W.~C. Mead, H.~C. Vantine, {\it Phys. Rev.
  Lett.\/} {\bf 35}, 1566 (1975).

\bibitem{Gonzalez-Alonso:2019a}
M.~Gonzalez-Alonso, O.~Naviliat-Cuncic, N.~Severijns, {\it Progress in Particle
  and Nuclear Physics\/} {\bf 104}, 165 (2019).

\bibitem{Holstein:2014a}
B.~Holstein, {\it J. Phys. G: Nucl. Part. Phys.\/} {\bf 41} (2014).

\bibitem{NavBSM2013}
O.~Naviliat-Cuncic, M.~González-Alonso, {\it Annalen der Physik\/} {\bf 525},
  600 (2013).

\bibitem{Cirigliano:2013xha}
V.~Cirigliano, S.~Gardner, B.~Holstein, {\it Prog. Part. Nucl. Phys.\/} {\bf
  71}, 93 (2013).

\bibitem{gardner13a}
S.~Gardner, B.~Plaster, {\it Phys. Rev. C\/} {\bf 87}, 065504 (2013).

\bibitem{Hardy:2015a}
J.~C. Hardy, I.~S. Towner, {\it Phys. Rev. C\/} {\bf 91}, 025501 (2015).

\bibitem{PDG:2019a}
M.~Tanabashi, {\it et~al.\/}, {\it Phys. Rev. D\/} {\bf 98}, 030001 (2018).

\bibitem{Cirigliano:2009wk}
V.~Cirigliano, J.~Jenkins, M.~Gonzalez-Alonso, {\it Nucl. Phys.\/} {\bf B830},
  95 (2010).

\bibitem{Bhattacharya:2011qm}
T.~Bhattacharya, {\it et~al.\/}, {\it Phys. Rev.\/} {\bf D85}, 054512 (2012).

\bibitem{Seng:2018a}
C.-Y. Seng, M.~Gorchtein, H.~H. Patel, M.~J. Ramsey-Musolf, {\it {Physical
  Review Letters}\/} {\bf {121}} ({2018}).

\bibitem{Seng:2019a}
C.-Y. Seng, M.~Gorchtein, M.~J. Ramsey-Musolf, {\it Phys. Rev. D\/} {\bf 100},
  013001 (2019).

\bibitem{Czarnecki:2019a}
A.~Czarnecki, W.~J. Marciano, A.~Sirlin, {\it Physical Review D\/} {\bf 100}
  (2019).

\bibitem{Towner:2010bx}
I.~S. Towner, J.~C. Hardy, {\it Reports on Progress in Physics\/} {\bf 73},
  046301 (2010).

\bibitem{NavSevHalfPlus2009}
O.~Naviliat-Cuncic, N.~Severijns, {\it Phys. Rev. Lett.\/} {\bf 102}, 142302
  (2009).

\bibitem{SevTownFtVals2008}
N.~Severijns, M.~Tandecki, T.~Phalet, I.~S. Towner, {\it Phys. Rev. C\/} {\bf
  78}, 055501 (2008).

\bibitem{Allen59}
J.~S. Allen, R.~L. Burman, W.~B. Hermannsfeldt, P.~Stahelin, T.~H. Braid, {\it
  Phys. Rev.\/} {\bf 116}, 134 (1959).

\bibitem{Calaprice69}
F.~P. Calaprice, E.~D. Commins, H.~M. Gibbs, G.~L. Wick, D.~A. Dobson, {\it
  Phys. Rev.\/} {\bf 184}, 1117 (1969).

\bibitem{Adelberger:1983}
E.~Adelberger, {\it et~al.\/}, {\it Physical Review C\/} {\bf 27}, 2833 (1983).

\bibitem{Hallin:1985}
A.~L. Hallin, {\it et~al.\/}, {\it Phys. Rev. Lett.\/} {\bf 52}, 337 (1984).

\bibitem{Lienard:2015a}
E.~Lienard, {\it et~al.\/}, {\it Hyperfine Interactions\/} {\bf 236}, 1 (2015).
  6th International Conference on Trapped Charged Particles and Fundamental
  Physics (TCP), Takamatsu, JAPAN, DEC 01-05, 2014.

\bibitem{Rebeiro:2019a}
B.~M. Rebeiro, {\it et~al.\/}, {\it {Physical Review C}\/} {\bf {99}} ({2019}).

\bibitem{Jackson1957}
J.~D. Jackson, S.~Treiman, H.~Wyld, {\it Phys. Rev.\/} {\bf 106}, 517 (1957).

\bibitem{NaviliatCuncic:2009gi}
O.~Naviliat-Cuncic, N.~Severijns, {\it Phys. Rev. Lett.\/} {\bf 102}, 142302
  (2009).

\bibitem{Severijns:2008gw}
N.~Severijns, M.~Tandecki, T.~Phalet, I.~Towner, {\it Phys. Rev. C\/} {\bf 78},
  055501 (2008).

\bibitem{Jones:1996}
G.~Jones, {A Measurement of the Beta Decay Asymmetry of 19Ne as a Test of the
  Standard Model}, Ph.D. thesis, Princeton University, Princeton University
  (1996).

\bibitem{Dobson:1963}
D.~A. Dobson, The beta-decay asymmetry and nuclear magnetic moment of neon-19,
  Ph.D. thesis, University of California Radiation Laboratory (1963).

\bibitem{Calaprice:1985}
F.~P. Calaprice, {\it Hyperfine Interactions\/} {\bf 22}, 83 (1985).

\bibitem{Calaprice:1991}
F.~P. Calaprice, W.~S. Anderson, G.~L. Jones, A.~R. Young, {\it AIP Conference
  Proceedings\/} {\bf 270}, 153 (1991).

\bibitem{Ramsey:1956}
N.~F. Ramsey, {\it Molecular Beams\/} (Oxford University Press, 1956).

\bibitem{Bielajew:2001jo}
A.~F. Bielajew, F.~Salvat, {\it Nuclear Instruments and Methods in Physics
  Research Section B: Beam Interactions with Materials and Atoms\/} {\bf 173},
  332 (2001).

\bibitem{Salvat:1995}
J.~Baro, J.~Sempau, J.~FernÃ¡ndez-Varea, F.~Salvat, {\it Nuclear Instruments
  and Methods in Physics Research Section B: Beam Interactions with Materials
  and Atoms\/} {\bf 100}, 31  (1995).

\bibitem{Knoll:2000}
G.~F. Knoll, {\it Radiation Detection and Measurement\/} (Wiley, New York,
  2000), third edn.

\bibitem{Spieler:1982}
H.~{Spieler}, {\it IEEE Transactions on Nuclear Science\/} {\bf 29}, 1142
  (1982).

\bibitem{behrens1982electron}
H.~Behrens, W.~B{\"u}hring, {\it Electron radial wave functions and nuclear
  betadecay\/}, no.~67 (Oxford University Press, USA, 1982).

\bibitem{Holstein:1974cw}
B.~Holstein, {\it Rev. Mod. Phys.\/} {\bf 46}, 789 (1974).

\bibitem{Hayen:2020a}
L.~Hayen, A.~R. Young, High precision predictions for angular correlations in
  allowed decays (2020). \url{https://arxiv.org/abs/2009.11364}.

\bibitem{Martin:2003a}
J.~W. Martin, {\it et~al.\/}, {\it Phys. Rev. C\/} {\bf 68}, 055503 (2003).

\bibitem{Martin:2006a}
J.~W. Martin, {\it et~al.\/}, {\it Phys. Rev. C\/} {\bf 73}, 015501 (2006).

\bibitem{plaster12}
B.~Plaster, {\it et~al.\/}, {\it Phys. Rev. C\/} {\bf 86}, 055501 (2012).

\bibitem{Schreiber:1983}
D.~F. Schreiber, {Beta Asymmetry of 19Ne: An Experimental Test for Second Class
  Currents}, Ph.D. thesis, Princeton University, Princeton University (1983).

\bibitem{heil2017}
W.~Heil, Measurements of he relaxation on mylar (2017).

\bibitem{jacob2003fundamental}
R.~Jacob, B.~Driehuys, B.~Saam, {\it Chemical physics letters\/} {\bf 370}, 261
  (2003).

\bibitem{cain1990nuclear}
E.~Cain, {\it et~al.\/}, {\it Journal of Physical Chemistry\/} {\bf 94}, 2128
  (1990).

\bibitem{Amsler:2008kq}
C.~Amsler, {\it et~al.\/}, {\it Physics Letters B\/} {\bf 667}, 1 (2008).

\bibitem{ame2016}
M.~Wang, {\it et~al.\/}, {\it Chinese Physics C\/} {\bf 41}, 030003 (2017).

\bibitem{towner2015parametrization}
I.~Towner, J.~Hardy, {\it Physical Review C\/} {\bf 91}, 015501 (2015).

\bibitem{Hayen2020c}
L.~Hayen pp. 1--23 (2020).

\bibitem{Bambynek1977}
W.~Bambynek, {\it et~al.\/}, {\it Reviews of Modern Physics\/} {\bf 49}, 77
  (1977).

\bibitem{Broussard:2014a}
L.~J. Broussard, {\it et~al.\/}, {\it Phys. Rev. Lett.\/} {\bf 112}, 212301
  (2014).

\bibitem{Fontbonne:2017mwx}
C.~Fontbonne, {\it et~al.\/}, {\it Phys. Rev. C\/} {\bf 96}, 065501 (2017).

\bibitem{Stone:2005ez}
N.~J. Stone, {\it Atomic Data and Nuclear Data Tables\/} {\bf 90}, 75 (2005).

\bibitem{Mohr:2012dr}
P.~J. Mohr, B.~N. Taylor, D.~B. Newell, {\it Journal of Physical and Chemical
  Reference Data\/} {\bf 41}, 043109 (2012).

\bibitem{adelberger1985}
E.~Adelberger, P.~Fernandez, C.~Gossett, J.~Osborne, V.~Zeps, {\it Physical
  review letters\/} {\bf 55}, 2129 (1985).

\bibitem{Marciano:2005ec}
W.~J. Marciano, A.~Sirlin, {\it Phys. Rev. Lett.\/} {\bf 96}, 032002 (2006).

\bibitem{Hayen:2019a}
L.~Hayen, N.~Severijns, Radiative corrections to gamow-teller decays (2019).
  \url{https://arxiv.org/abs/1906.09870}.

\bibitem{Gorchtein2018}
M.~Gorchtein, {\it Physical Review Letters\/} {\bf 123}, 042503 (2019).

\bibitem{Profumo:2007a}
S.~Profumo, M.~J. Ramsey-Musolf, S.~Tulin, {\it Phys. Rev. D\/} {\bf 75},
  075017 (2007).

\bibitem{Gonzalez-Alonso:2016a}
M.~Gonzalez-Alonso, O.~Naviliat-Cuncic, {\it Phys. Rev. C.\/} {\bf 94}, 035503
  (2016).

\bibitem{Gupta:2018a}
R.~Gupta, {\it et~al.\/}, {\it {Physical Review D}\/} {\bf {98}} ({2018}).

\bibitem{Pattie:2013a}
R.~Pattie~Jr., K.~Hickerson, A.~Young, {\it Phys. Rev. C\/} {\bf 88}, 048501
  (2013).

\bibitem{Pattie:2013b}
R.~W. Pattie, K.~P. Hickerson, A.~R. Young, {\it Phys. Rev. C\/} {\bf 92},
  069902 (2015).

\bibitem{plasterva2013}
S.~Gardner, B.~Plaster, {\it Phys. Rev. C\/} {\bf 87}, 065504 (2013).

\bibitem{Gysbers2019}
P.~Gysbers, {\it et~al.\/}, {\it Nature Physics\/}  (2019).

\bibitem{Barrett2013}
B.~R. Barrett, P.~Navr{\'{a}}til, J.~P. Vary, {\it Progress in Particle and
  Nuclear Physics\/} {\bf 69}, 131 (2013).

\bibitem{Fenker:2016a}
B.~Fenker, {\it et~al.\/}, {\it New Journal of Physics\/} {\bf 18}, 073028
  (2016).

\bibitem{Fenker:2017a}
B.~Fenker, {\it et~al.\/}, {\it {Physical Review Letters}\/} {\bf {120}}
  ({2018}).

\bibitem{Melconian:2019a}
D.~Melconian, J.~Behr, unpublished (2019).

\bibitem{Mardor:2018a}
I.~Mardor, {\it et~al.\/}, {\it The European Physical Journal A\/} {\bf 54}, 91
  (2018).

\end{thebibliography}

\bibliographystyle{Science}

\section*{Acknowledgments}
  ARY is grateful for the sustained encouragement and enthusiasm of Stuart Freedman concerning his involvement in fundamental symmetries research and this result in particular. The efforts of undergraduate student A. Ackerson were critical in characterizing the Si(Li) geometry and establishing realistic values for the operating temperature.  We also want to thank P. Ciaccio for his help with the atomic beam figure. The experiment would not have been possible without the support and daily assistance of Fred Loeser and Amir Razzaghi at the Princeton Cyclotron. This publication benefited greatly from conversations with Leah Broussard, Vincenzo Cirigliano, Stuart Freedman, Alejandro Garcia, John Hardy, Mikhail Gorchstein, Michael Ramsey-Musolf and Ian Towner.  The research was supported by NSF grants: 9015631, 9312588, 0653222, 1005233, 1307426, 1615153, 1914133 and DOE grant DE-FG02-ER41042).


\clearpage

\end{document}